\documentclass[12pt,a4paper]{article}

\usepackage{ifthen} 
\newboolean{pdflatex}
\setboolean{pdflatex}{true} 

\usepackage{subfigure}

\newboolean{articletitles}
\setboolean{articletitles}{true} 

\newboolean{uprightparticles}
\setboolean{uprightparticles}{false} 

\newboolean{inbibliography}
\setboolean{inbibliography}{false} 

\usepackage[top=1in, bottom=1.25in, left=1in, right=1in]{geometry}

\usepackage{microtype}
\usepackage{lineno}  
\usepackage{xspace} 
\usepackage{caption} 

\usepackage{graphicx}  
\usepackage{color}
\usepackage{colortbl}
\graphicspath{{./figs/}} 

\usepackage{amsmath} 
\usepackage{amssymb}
\usepackage{amsfonts}
\usepackage{upgreek} 

\newcommand*\patchAmsMathEnvironmentForLineno[1]{%
\expandafter\let\csname old#1\expandafter\endcsname\csname #1\endcsname
\expandafter\let\csname oldend#1\expandafter\endcsname\csname
end#1\endcsname
 \renewenvironment{#1}%
   {\linenomath\csname old#1\endcsname}%
   {\csname oldend#1\endcsname\endlinenomath}%
}
\newcommand*\patchBothAmsMathEnvironmentsForLineno[1]{%
  \patchAmsMathEnvironmentForLineno{#1}%
  \patchAmsMathEnvironmentForLineno{#1*}%
}
\AtBeginDocument{%
\patchBothAmsMathEnvironmentsForLineno{equation}%
\patchBothAmsMathEnvironmentsForLineno{align}%
\patchBothAmsMathEnvironmentsForLineno{flalign}%
\patchBothAmsMathEnvironmentsForLineno{alignat}%
\patchBothAmsMathEnvironmentsForLineno{gather}%
\patchBothAmsMathEnvironmentsForLineno{multline}%
\patchBothAmsMathEnvironmentsForLineno{eqnarray}%
}

\usepackage{hyperref}    
\usepackage[all]{hypcap} 

\usepackage{xspace} 
\usepackage{upgreek}

\def\lhcb {\mbox{LHCb}\xspace}

\def\MagUp {\mbox{\em Mag\kern -0.05em Up}\xspace}

\ifthenelse{\boolean{uprightparticles}}%
{

 \def\Ppi         {\ensuremath{\uppi}\xspace}

 \def\Ppsi        {\ensuremath{\uppsi}\xspace}

 \def\PDelta      {\ensuremath{\Delta}\xspace}                 
 \def\PXi      {\ensuremath{\Xi}\xspace}                 
 \def\PLambda      {\ensuremath{\Lambda}\xspace}                 
 \def\PSigma      {\ensuremath{\Sigma}\xspace}                 
 \def\POmega      {\ensuremath{\Omega}\xspace}                 
 \def\PUpsilon      {\ensuremath{\Upsilon}\xspace}                 
 
 \def\PB      {\ensuremath{\mathrm{B}}\xspace}                 
                  
 \def\PD      {\ensuremath{\mathrm{D}}\xspace}

 \def\PJ      {\ensuremath{\mathrm{J}}\xspace}                 
 \def\PK      {\ensuremath{\mathrm{K}}\xspace}

 \def\PW      {\ensuremath{\mathrm{W}}\xspace}

 \def\Pc      {\ensuremath{\mathrm{c}}\xspace}

 \def\Pi      {\ensuremath{\mathrm{i}}\xspace}

 \def\Ps      {\ensuremath{\mathrm{s}}\xspace}

}
{

 \def\Ppi         {\ensuremath{\pi}\xspace}

 \def\Ppsi        {\ensuremath{\psi}\xspace}                 
                  
 \mathchardef\PDelta="7101
 \mathchardef\PXi="7104
 \mathchardef\PLambda="7103
 \mathchardef\PSigma="7106
 \mathchardef\POmega="710A
 \mathchardef\PUpsilon="7107
                  
 \def\PB      {\ensuremath{B}\xspace}                 
                  
 \def\PD      {\ensuremath{D}\xspace}

 \def\PJ      {\ensuremath{J}\xspace}                 
 \def\PK      {\ensuremath{K}\xspace}

 \def\PW      {\ensuremath{W}\xspace}

 \def\Pc      {\ensuremath{c}\xspace}

 \def\Pi      {\ensuremath{i}\xspace}

 \def\Ps      {\ensuremath{s}\xspace}

}

\makeatletter
\ifcase \@ptsize \relax
  \newcommand{\miniscule}{\@setfontsize\miniscule{4}{5}}
\or
  \newcommand{\miniscule}{\@setfontsize\miniscule{5}{6}}
\or
  \newcommand{\miniscule}{\@setfontsize\miniscule{5}{6}}
\fi
\makeatother

\DeclareRobustCommand{\optbar}[1]{\shortstack{{\miniscule (\rule[.5ex]{1.25em}{.18mm})}
  \\ [-.7ex] $#1$}}

\def\Wp     {{\ensuremath{\PW^+}}\xspace}

\def\squark    {{\ensuremath{\Ps}}\xspace}

\def\cquark    {{\ensuremath{\Pc}}\xspace}

\def\pion   {{\ensuremath{\Ppi}}\xspace}
\def\piz    {{\ensuremath{\pion^0}}\xspace}

\def\pip    {{\ensuremath{\pion^+}}\xspace}
\def\pim    {{\ensuremath{\pion^-}}\xspace}

\def\kaon    {{\ensuremath{\PK}}\xspace}
  \def\Kbar    {{\kern 0.2em\overline{\kern -0.2em \PK}{}}\xspace}

\def\KorKbar    {\kern 0.18em\optbar{\kern -0.18em K}{}\xspace}

\def\Kp      {{\ensuremath{\kaon^+}}\xspace}
\def\Km      {{\ensuremath{\kaon^-}}\xspace}

  \def\Dbar    {{\kern 0.2em\overline{\kern -0.2em \PD}{}}\xspace}
\def\D       {{\ensuremath{\PD}}\xspace}

\def\DorDbar    {\kern 0.18em\optbar{\kern -0.18em D}{}\xspace}
\def\Dz      {{\ensuremath{\D^0}}\xspace}
\def\Dzb     {{\ensuremath{\Dbar{}^0}}\xspace}

\def\Dstarz  {{\ensuremath{\D^{*0}}}\xspace}

\def\Dstarp  {{\ensuremath{\D^{*+}}}\xspace}

\def\B       {{\ensuremath{\PB}}\xspace}
\def\Bbar    {{\ensuremath{\kern 0.18em\overline{\kern -0.18em \PB}{}}}\xspace}

\def\BorBbar    {\kern 0.18em\optbar{\kern -0.18em B}{}\xspace}
\def\Bz      {{\ensuremath{\B^0}}\xspace}

\def\Bu      {{\ensuremath{\B^+}}\xspace}

\def\Bp      {{\ensuremath{\Bu}}\xspace}

\def\Bs      {{\ensuremath{\B^0_\squark}}\xspace}

\def\Bc      {{\ensuremath{\B_\cquark^+}}\xspace}
\def\Bcp     {{\ensuremath{\B_\cquark^+}}\xspace}

\def\jpsi     {{\ensuremath{{\PJ\mskip -3mu/\mskip -2mu\Ppsi\mskip 2mu}}}\xspace}

  \def\Y#1S{\ensuremath{\PUpsilon{(#1S)}}\xspace}

\def\Lbar        {{\ensuremath{\kern 0.1em\overline{\kern -0.1em\PLambda}}}\xspace}
\def\LorLbar    {\kern 0.18em\optbar{\kern -0.18em \PLambda}{}\xspace}

\def\to                 {\ensuremath{\rightarrow}\xspace}

\def\AT#1     {\ensuremath{A_{\mathrm{T}}^{#1}}\xspace}           

\def\C#1      {\ensuremath{\mathcal{C}_{#1}}\xspace}                       
\def\Cp#1     {\ensuremath{\mathcal{C}_{#1}^{'}}\xspace}                    
\def\Ceff#1   {\ensuremath{\mathcal{C}_{#1}^{\mathrm{(eff)}}}\xspace}        
\def\Cpeff#1  {\ensuremath{\mathcal{C}_{#1}^{'\mathrm{(eff)}}}\xspace}       
\def\Ope#1    {\ensuremath{\mathcal{O}_{#1}}\xspace}                       
\def\Opep#1   {\ensuremath{\mathcal{O}_{#1}^{'}}\xspace}                    

\newcommand{\tev}{\ifthenelse{\boolean{inbibliography}}{\ensuremath{~T\kern -0.05em eV}\xspace}{\ensuremath{\mathrm{\,Te\kern -0.1em V}}}\xspace}
\newcommand{\gev}{\ensuremath{\mathrm{\,Ge\kern -0.1em V}}\xspace}
\newcommand{\mev}{\ensuremath{\mathrm{\,Me\kern -0.1em V}}\xspace}
\newcommand{\kev}{\ensuremath{\mathrm{\,ke\kern -0.1em V}}\xspace}
\newcommand{\ev}{\ensuremath{\mathrm{\,e\kern -0.1em V}}\xspace}
\newcommand{\gevc}{\ensuremath{{\mathrm{\,Ge\kern -0.1em V\!/}c}}\xspace}
\newcommand{\mevc}{\ensuremath{{\mathrm{\,Me\kern -0.1em V\!/}c}}\xspace}
\newcommand{\gevcc}{\ensuremath{{\mathrm{\,Ge\kern -0.1em V\!/}c^2}}\xspace}
\newcommand{\gevgevcccc}{\ensuremath{{\mathrm{\,Ge\kern -0.1em V^2\!/}c^4}}\xspace}
\newcommand{\mevcc}{\ensuremath{{\mathrm{\,Me\kern -0.1em V\!/}c^2}}\xspace}

\def\invfb   {\ensuremath{\mbox{\,fb}^{-1}}\xspace}

\def\gsim{{~\raise.15em\hbox{$>$}\kern-.85em
          \lower.35em\hbox{$\sim$}~}\xspace}
\def\lsim{{~\raise.15em\hbox{$<$}\kern-.85em
          \lower.35em\hbox{$\sim$}~}\xspace}

\def\pt         {\mbox{$p_{\mathrm{ T}}$}\xspace}

\def\tell1  {TELL1\xspace}
\def\ukl1   {UKL1\xspace}

\usepackage{cite} 
\usepackage{mciteplus}

\usepackage{longtable} 
\usepackage{tikz}
\usetikzlibrary{decorations.pathreplacing,decorations.pathmorphing,decorations.markings,trees,calc,patterns,snakes,arrows}
\tikzset{
photon/.style={decorate, decoration={snake}, draw=red},
particle/.style={draw=blue, postaction={decorate},decoration={markings,mark=at position .5 with {\arrow[draw=blue]{>}}}},
antiparticle/.style={draw=blue, postaction={decorate},decoration={markings,mark=at position .5 with {\arrow[draw=blue]{<}}}}, 
gluon/.style={decorate, draw=black,decoration={coil,amplitude=4pt, segment length=5pt}}, 
majorana/.style={draw=black, postaction={decorate},decoration={markings,mark=at position .48 with {\arrow[draw=black]{>}},mark=at position .52 with {\arrow[draw=black]{<}}}},
gluonloop/.style={circle, decorate, draw=black, decoration={coil,aspect=1.2,amplitude=2pt, segment length=4pt},minimum height=1.2em},
}

\begin{document}

\renewcommand{\thefootnote}{\fnsymbol{footnote}}
\setcounter{footnote}{1}



\begin{titlepage}
\pagenumbering{roman}

\vspace*{-1.5cm}
\centerline{\large EUROPEAN ORGANIZATION FOR NUCLEAR RESEARCH (CERN)}
\vspace*{1.5cm}
\noindent
\begin{tabular*}{\linewidth}{lc@{\extracolsep{\fill}}r@{\extracolsep{0pt}}}
\ifthenelse{\boolean{pdflatex}}
{\vspace*{-2.7cm}\mbox{\!\!\!\includegraphics[width=.14\textwidth]{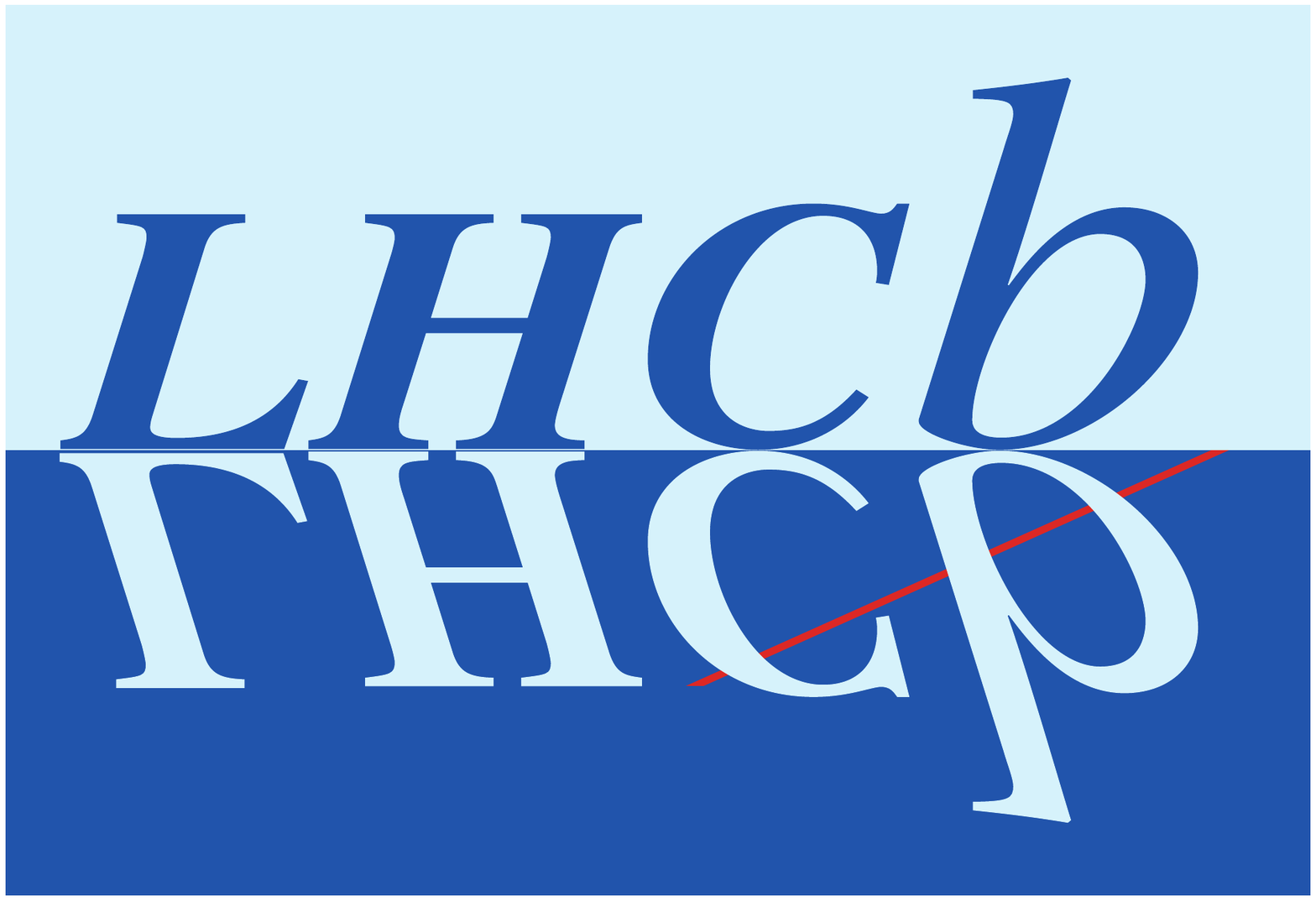}} & &}%
{\vspace*{-1.2cm}\mbox{\!\!\!\includegraphics[width=.12\textwidth]{lhcb-logo.eps}} & &}%
\\
 & & CERN-EP-2016-314 \\  
 & & LHCb-PAPER-2016-058 \\  
 & & March 15, 2017 \\ 
 & & \\
\end{tabular*}

\vspace*{1.0cm}

{\normalfont\bfseries\boldmath\huge
\begin{center}
  Observation of $\Bcp\to\Dz\Kp$ decays
\end{center}
}

\vspace*{1.0cm}

\begin{center}
The LHCb collaboration\footnote{Authors are listed at the end of this letter.}
\end{center}

\vspace*{1.0cm}

\begin{abstract}
  \noindent
  Using proton-proton collision data corresponding to an integrated luminosity of 3.0\invfb, recorded by the LHCb detector at
centre-of-mass energies of 7 and 8\tev, the $\Bcp\to\Dz\Kp$ decay is observed with a statistical significance of 5.1 standard deviations.
  By normalising to $\Bp\to\Dzb\pip$ decays, a measurement of the branching fraction multiplied by the production rates for \Bcp relative to \Bp mesons in the LHCb acceptance is obtained,
$$
R_{\Dz K} = \frac{f_{c}}{f_{u}}\times\mathcal{B}(\Bcp\to\Dz\Kp) = (9.3\,^{+2.8}_{-2.5} \pm 0.6) \times 10^{-7}\,,
$$
where the first uncertainty is statistical and the second is systematic. This decay is expected to proceed predominantly through weak annihilation and penguin amplitudes, and is the first \Bcp decay of this nature to be observed.

\end{abstract}
\vspace*{2.0cm}

\begin{center}
  Published in Phys.~Rev.~Lett \textbf{118}, 111803 (2017)
\end{center}

\vspace{\fill}

{\footnotesize 
\centerline{\copyright~CERN on behalf of the \lhcb collaboration, licence \href{http://creativecommons.org/licenses/by/4.0/}{CC-BY-4.0}.}}
\vspace*{2mm}

\end{titlepage}


\newpage
\setcounter{page}{2}
\mbox{~}
%
%
%
%

\cleardoublepage


\renewcommand{\thefootnote}{\arabic{footnote}}
\setcounter{footnote}{0}



\pagestyle{plain} 
\setcounter{page}{1}
\pagenumbering{arabic}



The \Bc meson is the only ground-state meson consisting of two heavy quarks of different flavour, namely a $\bar{b}$ and a $c$ quark. 
As such, its formation in $pp$ collisions is suppressed relative to the lighter \B mesons. 
Unlike \Bz, \Bp and \Bs mesons, the $b$-quark decay accounts for only $\sim20\%$ of the \Bc width~\cite{Gouz:2002kk}.
Around 70\% of its width is due to $c$-quark decays, where the $c$-quark transition has been observed with \mbox{$\Bc\to\Bs\pip$} decays~\cite{LHCb-PAPER-2013-044}. This leaves $\sim10\%$ for $\bar b c\to\Wp\!\to \bar q q$ annihilation amplitudes, which can be unambiguously probed in charmless final states. No charmless \Bc decays have been reported to date, although searches show an indication at the level of $2.4$ standard deviations ($\sigma$)~\cite{Aaij:2016xas}. 

\begin{figure}[!b]
\begin{center}
  \includegraphics*[width=1.00\textwidth]{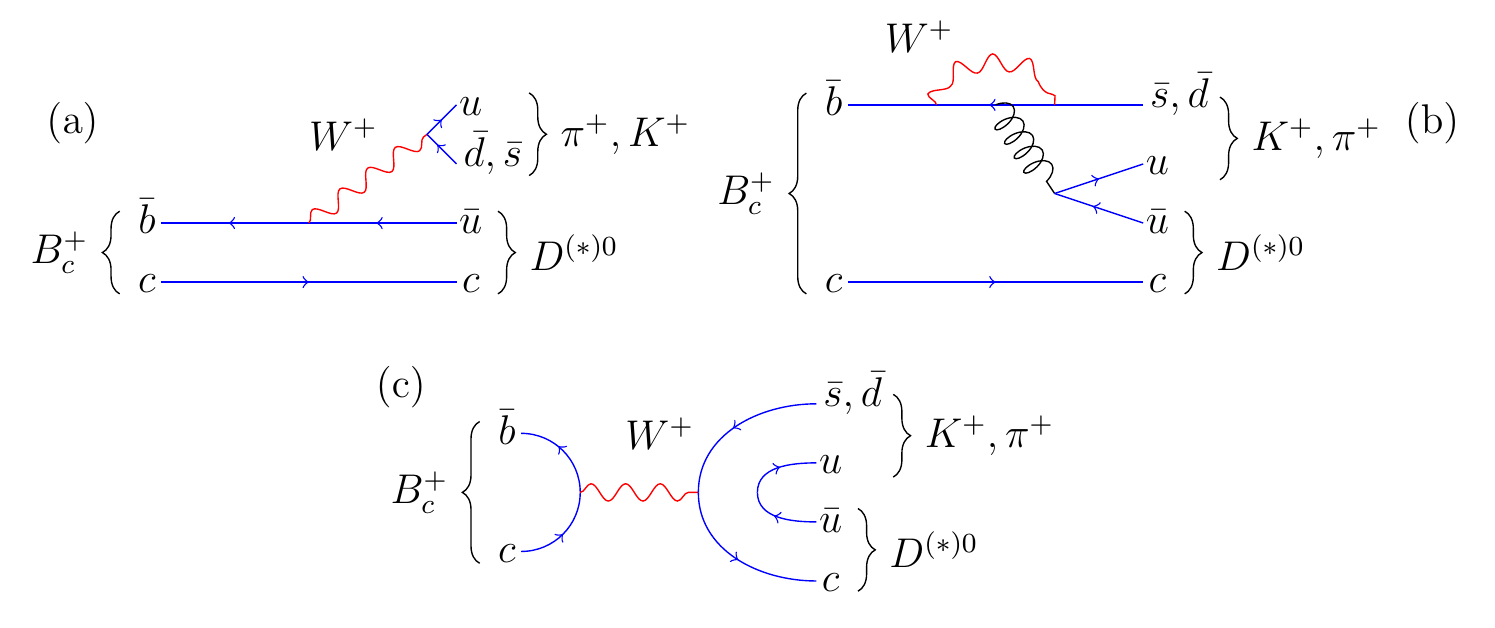}
    \caption{Tree (a), penguin (b), and weak annihilation (c) diagrams for the decays studied. In each case, the meson appearing before the comma denotes the favoured decay. 
  \label{fig:feyn}}
  \end{center}
\end{figure}

To test QCD factorisation and explore the new physics potential of \Bc decays, rarer decays such as suppressed tree-level $b\to u$ transitions and $b\to s$ loop-mediated (penguin) decays can be studied, where the charm quantum number remains unchanged. The simplest decay is the colour-allowed $\Bc\to D^{(*)0}\pip$ decay, illustrated in Fig.\,\ref{fig:feyn}(a).
The expected branching fraction for this decay is a factor $|V_{ub}/V_{cb}|^2\approx0.007$ lower than the favoured $b \to c$ and colour-allowed $\Bc\to \jpsi \pip$ decay~\cite{Abazov:2008kv,LHCb-PAPER-2014-050}, placing this mode at the limit of sensitivity with current LHCb data. However, this expectation may be enhanced by penguin and weak annihilation amplitudes, which will be more pronounced in the \mbox{$\Bc\to D^{(*)0} \Kp$} mode (see Fig.\,\ref{fig:feyn}(b,c)). This motivates a search for the \mbox{$\Bcp \to D^{(*)0}\Kp$ and $\Bcp \to D^{(*)0}\pip$} decays, particularly as the branching fraction estimates in the literature vary considerably~\cite{Bc_theory,Bc_theory3,Zhang2009}. 

The decay $\Bp \to \Dzb \pip$ is used for normalisation. Since the ratio of production rates for \Bc and \Bu mesons within the LHCb acceptance, $f_{c}/f_{u}$, is unknown, the measured observables are
\begin{align}
R_{D^{(*)0}h} = \frac{f_{c}}{f_{u}} \times \mathcal{B}(\Bcp \to D^{(*)0} h^{+})\,, \label{eq1}
\end{align}
where $h$ is $\pi$ or $K$ and $\mathcal{B}(\Bcp \to \D^{(*)0} h^{+})$ represents the corresponding branching fraction. The four observables are measured with a simultaneous fit to the $\Dz\pip$ and $\Dz\Kp$ invariant mass distributions. Theoretical estimates for \mbox{$\mathcal{B}(\Bcp \to \jpsi \pip)$} range from \mbox{$6.0 \times 10^{-4}$}~\cite{PhysRevD.68.094020} to \mbox{$1.8 \times 10^{-3}$}~\cite{PhysRevD.49.3399}, which implies $f_{c}/f_{u}$ values in the range $0.004 - 0.012$ using the production ratio measured in Ref.~\cite{LHCb-PAPER-2014-050} and the branching fraction $\mathcal{B}(\Bp \to J/\psi \Kp)$~\cite{Olive:2016xmw}. Estimates for \mbox{$\mathcal{B}(\Bcp \to \Dz \Kp)$} vary from $1.3 \times 10^{-7}$~\cite{Bc_theory} to $6.6 \times 10^{-5}$~\cite{Zhang2009}, while estimates for \mbox{$\mathcal{B}(\Bcp \to \Dz \pip)$} vary from $2.3 \times 10^{-7}$~\cite{Bc_theory} to $2.3 \times 10^{-6}$~\cite{Bc_theory3}. Using Eq.~(\ref{eq1}), the expectation for $R_{\Dz\pi}$ is seen to cover the range $9 \times 10^{-10} - 3 \times 10^{-8}$, while $R_{\Dz K}$ covers the range $5 \times 10^{-10} - 8 \times 10^{-7}$.

This Letter reports a search for $\Bc \to \Dz \pip$ and $\Bc \to \Dz \Kp$ decays in $pp$ collision data corresponding to integrated luminosities of 1.0 and 2.0\invfb taken by the LHCb experiment at centre-of-mass energies of 7 and 8\tev, respectively, where the \Dz meson is reconstructed in the Cabibbo-favoured final states $\Dz \to \Km\pip$ or $\Dz \to \Km \pip \pim \pip$ (inclusion of charge-conjugate processes is implied throughout). Partially reconstructed $\Bcp\to (\Dstarz \to \Dz \{\piz,\gamma\})h^+$ decays, where the neutral particle indicated in braces is not considered in the invariant mass calculation, are treated as additional signal channels. The number of $\Bcp$ decays is normalised by comparison to the number of \mbox{$\Bp \to (\Dzb \to \Kp \pim (\pip \pim)) \pip$} decays. A fit to the invariant mass distribution of $D^{(*)0}h^+$ candidates in the range \mbox{$5800 - 6900$ \mevcc} enables a measurement of
\begin{align}
R_{D^{(*)0}h} = \frac{\mathcal{N}(\Bcp \rightarrow D^{(*)0}h^{+})}{\mathcal{N}(\Bp \rightarrow \Dzb \pip)} \times \mathcal{B}(\Bp \rightarrow \Dzb \pip) \times \xi\,, \label{r_Bc_h_definition}
\end{align}
where $\mathcal{N}(\Bcp \rightarrow D^{(*)0}h^{+})$ represents the $\Bcp \to \D^{(*)0} h^{+}$ yield, $\mathcal{N}(\Bp \rightarrow \Dzb \pi^{+})$ represents the yield of $\Bp \to \Dzb \pip$ normalisation decays, $\mathcal{B}(\Bp \rightarrow \Dzb \pip)$ is the normalisation mode branching fraction~\cite{Olive:2016xmw}, and $\xi$ is the ratio of efficiencies for reconstructing and selecting $\Bp$ and $\Bcp$ mesons decaying to these final states. 


The \lhcb detector is a single-arm forward spectrometer covering the \mbox{pseudorapidity} range $2<\eta <5$, described in detail in Refs.~\cite{Alves:2008zz,LHCb-DP-2014-002}. The detector allows the reconstruction of both charged and neutral particles. For this analysis, the ring-imaging Cherenkov (RICH) detectors \cite{RichPerf}, distinguishing pions, kaons and protons, are particularly important. Simulated events are produced using the software described in Refs.~\cite{Sjostrand:2006za,Sjostrand:2007gs,LHCb-PROC-2010-056,Lange:2001uf, Allison:2006ve,Agostinelli:2002hh,LHCb-PROC-2011-006,Chang:2003cq}.


After reconstruction of the \Dz meson candidate, the same selection is applied to the \Bcp and \Bp candidates. The invariant mass of the \Dz candidate must be within $\pm 25 \mevcc$ of its known value~\cite{Olive:2016xmw}. The other hadron originating from the \B decay must have transverse momentum (\pt) in the range \mbox{$0.5 - 10.0$ \gevc} and momentum ($p$) in the range \mbox{$5 - 100$ \gevc}, ensuring that the track is within the kinematic coverage of the RICH detectors that provide particle identification (PID) information. A kinematic fit is performed to each decay chain~\cite{Hulsbergen:2005pu}, with vertex constraints applied to both the \B and \D vertices, and the \Dz candidate mass constrained to its known value.
The \Bp ($\Bcp$) meson candidates with an invariant mass in the interval \mbox{$5080 - 5900$ \mevcc} \mbox{($5800 - 6900$ \mevcc)} and with a proper decay time above 0.2 ps are retained.
Each \B candidate is associated to the primary vertex (PV) to which it has the smallest impact parameter (IP), defined as the distance of closest approach of the candidate's trajectory to a given PV. 
 
 Two boosted decision tree (BDT) discriminators~\cite{Roe} are used for further background suppression.
They are trained using simulated \mbox{$\Bcp \to (\Dz \to \Km \pip (\pip\pim))h^{+}$} signal decays and a sample of wrong-sign $\Kp\pim(\pip\pim)h^{+}$ combinations from data with invariant mass in the range \mbox{$5900-7200$ \mevcc}. 
For the first BDT, background candidates with a \Dz invariant mass more than $\pm 30$ \mevcc away from the known \Dz mass are used. In the second BDT, background candidates with a \Dz invariant mass within $\pm25$ \mevcc of the known \Dz mass are used.
A loose cut on the classifier response of the first BDT is applied before training the second one. This focusses the second BDT training on backgrounds enriched with fully reconstructed \Dz mesons. 

The inputs to all BDTs include properties of each particle ($p$, \pt, and the IP significance) and additional properties of the \B and \Dz candidates (decay time, flight distance, decay vertex quality, radial distance between the decay vertex and the PV, and the angle between the reconstructed momentum vector and the line connecting the production and decay vertices). A two-dimensional optimisation is performed to determine the second stage BDT requirements for the two-body and four-body modes, where the signal \textit{S} is compared to the number of background events \textit{B} in data using a figure of merit \mbox{$\textit{S}/(\sqrt{\textit{B}} + 3/2)$}~\cite{Punzi:2003bu}. The value of \textit{B} is determined within $\pm50$ \mevcc of the known \Bcp mass. No PID information is used in the BDT training, so that the efficiency for $\B \to \Dz \Kp$ and $B \to \Dz \pip$ decays is similar. The use of BDTs to select signal decays was validated by comparing the efficiency of the BDT requirements for $\Bp \to \Dzb \pip$ decays in data and simulation, where close agreement was found across a wide range of BDT cuts. The purity of the selection is further improved by requiring all kaons and pions in the \Dz decay to be identified with a PID selection that has an efficiency of about 85\% per particle. 

Simulated signal samples are used to evaluate the relative efficiency for selecting $\Bcp$ and $\Bp$ decays. 
The efficiency ratio is $\xi = \epsilon(\Bp)/\epsilon(\Bcp)$, where $\epsilon(\Bp)$ and $\epsilon(\Bcp)$ represent the combined 
efficiencies of detector acceptance, trigger, reconstruction and offline selection. As both \Bcp and \Bp mesons are required to decay to the same final-state particles, differences between $\epsilon(\Bp)$ and $\epsilon(\Bcp)$ arise due to differences in their masses and lifetimes. The \Bcp meson lifetime is $(0.507 \pm 0.009)$ ps, which is 3.2 times shorter than that of the \Bp meson~\cite{Olive:2016xmw}. This results in a lower \Bcp efficiency relative to \Bp by a factor $2.4$, due to the proper decay time cut. The \Bcp meson is heavier than the \Bp, which reduces by a factor $1.3$ the fraction of $\Bcp$ decays in which all final-state particles are within the detector acceptance. However, as the BDTs are trained specifically on \Bcp simulated decays, the offline selection efficiency is lower for \Bp decays, contributing a relative efficiency of $0.94$. Overall, the efficiency ratio is $\xi = 3.04 \pm 0.16$ $(2.88 \pm 0.15)$ for the two-body (four-body) \Dz decay. The uncertainties are systematic, arising from the use of finite simulated samples and possible mismodelling of the simulated \Bcp lifetime and production kinematics.



To measure $\mathcal{N}(\Bp \to \Dzb \pip)$, binned maximum likelihood fits to the invariant mass distributions of selected \Bp candidates are performed, where separate fits are employed for the two-body and four-body $\Dzb$ modes. The total probability density function (PDF) is built from four contributions. The $\Bp \to \Dzb \pip$ decays are modelled by the sum of two modified Gaussian functions with asymmetric power-law tails and an additional Gaussian function as used in Ref.~\cite{LHCb-PAPER-2016-003}, all of which share a common peak position. Misidentified $\Bp \to \Dzb \Kp$ candidates have an incorrect mass assignment and form a distribution displaced downward in mass, with a tail extending to lower invariant masses. They are modelled by the sum of two modified Gaussian PDFs with low-mass power-law tails. All PDF parameters are allowed to vary, with the exception of the tail parameters which are fixed to the values found in simulation. 

Partially reconstructed decays form a background at invariant masses lower than that of the signal peak. This background is described by a combination of parametric PDFs, with yield and shape parameters that are allowed to vary. A linear function describes the combinatorial background. The yield of $\Bp \to \Dzb \Kp$ decays, where the kaon is misidentified as a pion, is fixed using a simultaneous fit to correctly identified $\Bp \to \Dzb \Kp$ events. Using a data-driven analysis of approximately 20 million \Dstarp decays reconstructed as $\Dstarp \to \Dz \pip, \Dz \to \Km \pip$, the probability of kaon misidentification is determined to be 32\%. The invariant mass fits to $\Bp \to (\Dzb \to \Kp \pim) \pip$ and \mbox{$\Bp \to (\Dzb \to \Kp \pim \pip \pim) \pip$} decays determine a total observed yield \mbox{$\mathcal{N}(\Bp \to \Dzb \pip) = 309\,462 \pm 550$.}

To measure $\mathcal{N}(\Bcp \rightarrow D^{(*)0}h^+)$, a simultaneous invariant mass fit to the \mbox{$\Bcp \to \Dz \pip$} and \mbox{$\Bcp \to \Dz\Kp$} samples is performed in the region $5800 - 6900 \mevcc$. Two-body and four-body \D-decay candidates are included, where a Gaussian PDF describes the fully reconstructed $\Bcp$ signals. The mean of this Gaussian is fixed to the known $\Bcp$ mass~\cite{Olive:2016xmw}. The width of the $\Bcp \to \Dz \pip$ PDF is taken from a fit to suppressed \mbox{$\Bp \to (\Dzb \to \pip \Km) \pip$} decays, scaled up by a factor 1.3 to account for the difference in momenta of the decay products in $\Bcp \to \Dz\pip$ and $\Bp \to \Dzb \pip$ decays. The width of the $\Bcp \to \Dz\Kp$ peak is related to that of $\Bcp \to \Dz\pip$ decays by the ratio of the widths of the $\Bp \to \Dzb \Kp$ and $\Bp \to \Dzb \pip$ peaks found in the normalisation mode fits. Partially reconstructed $\Bcp \to \Dstarz h^+$ signal decays are modelled using a combination of parametric PDFs, with yield and shape parameters that are allowed to vary. These decays contribute at lower invariant masses than the fully reconstructed signal decays, as a result of not considering the natural particle in the invariant mass calculation. An additional background component at low invariant mass is included to describe $\Bcp$ decays where two particles are missed, with shape parameters taken from simulated $\Bp \to \Dstarz \pip \piz$ decays and scaled to account for the different momenta of the decay products in $\Bcp$ and $\Bp$ decays. 

\begin{figure}[ht]
  \begin{center}
    \includegraphics*[width=0.65\textwidth]{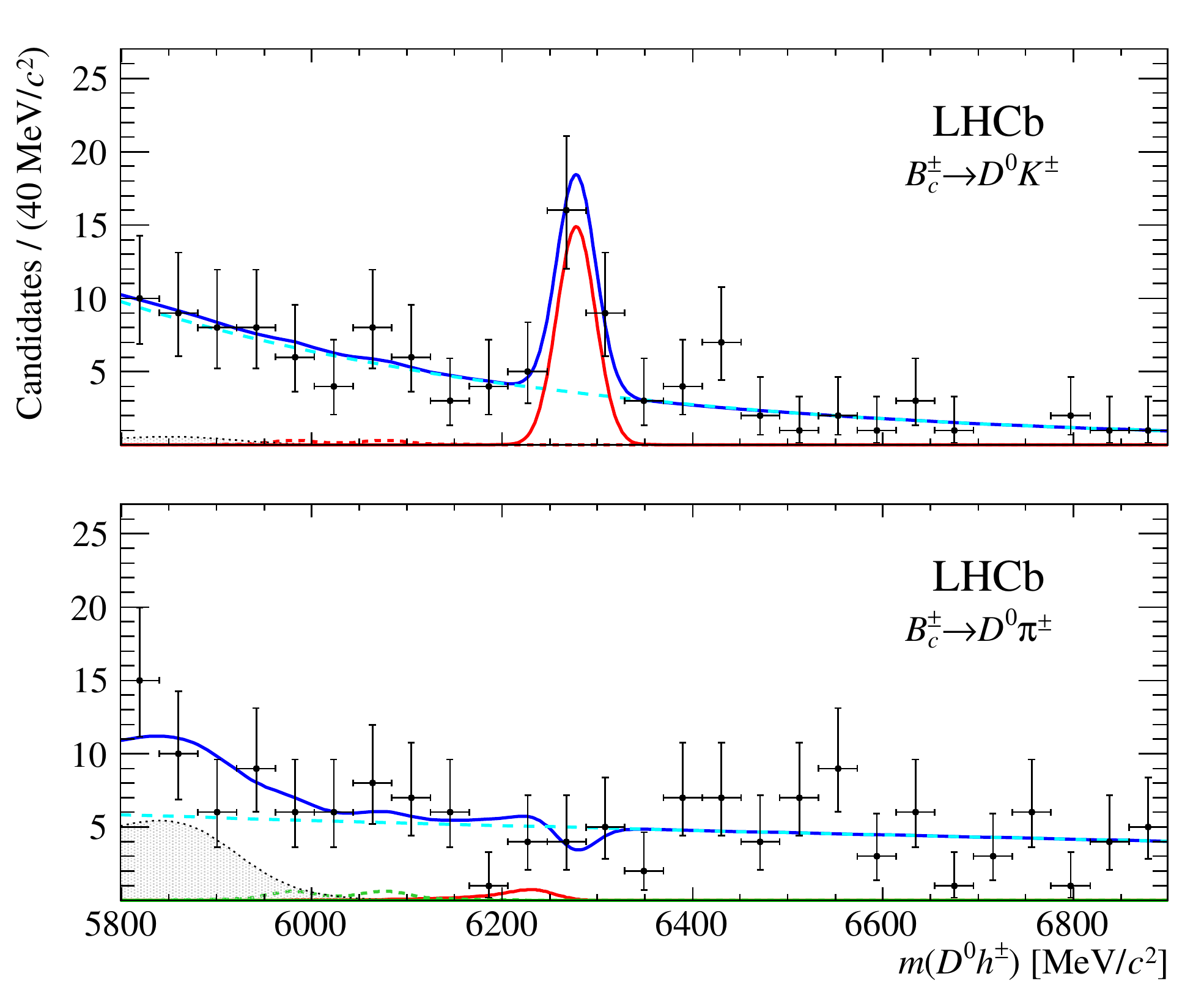}
  \caption{Results of the simultaneous fit to the $\Dz\Kp$ (top plot) and $\Dz\pip$ (bottom plot) invariant mass distributions in the $\Bcp$ mass region, including the $\Dz \rightarrow \Km \pip$ and \mbox{$\Dz \rightarrow \Km \pip \pim \pip$} final states.  Inclusion of the charge conjugate decays is implied. The red solid curve illustrates $\Bcp \to \Dz K^+$ decays, the red dashed curve illustrates $\Bcp \to \Dstarz K^+$ decays, the green dashed curve represents $\Bcp \to \Dstarz \pi^+$ decays, the grey shaded region represents partially reconstructed background decays, the cyan dashed line represents the combinatorial background, and the total PDF is displayed as a blue solid line. The small drop visible in the total $\Bcp \to D^{(*)0} \pip$ PDF around the \Bcp mass arises from the fact that the fit finds a small negative value for the $\Bcp \to \Dz\pip$ yield.
  \label{Bc_mass_fit}}
  \end{center}
  \end{figure}

Misidentified $\Bcp \to \Dz\pip(\Kp)$ decays in the $\Bcp \to \Dz\Kp(\pip)$ sample are modelled using the same PDFs as the normalisation fits, with widths and peak positions scaled for the decay momentum difference. These shapes are fixed in the fit. Signal decays are split into separate samples with correct and incorrect kaon identification, with a kaon misidentification rate of 7\% and a corresponding pion identification efficiency of 91\% fixed using the data-driven \Dstarp analysis described above. An exponential function describes the combinatorial background, which is fitted independently in the $\Bcp \to \Dz\pip$ and $\Bcp \to \Dz\Kp$ samples. The combinatorial yields, signal yields and partially reconstructed $\Bcp \to \Dz h^+ \{\piz\}$ and $\Bcp \to D^{*0} h^+ \{\piz\}$ background yields are all free to vary. The fit to data is shown in Fig.\,\ref{Bc_mass_fit}, where a $\Bcp \to \Dz\Kp$ yield of $20 \pm 5$ events is found. All other signal yields are consistent with zero. 

\begin{figure}
  \begin{center}
    \includegraphics[width=0.4\linewidth]{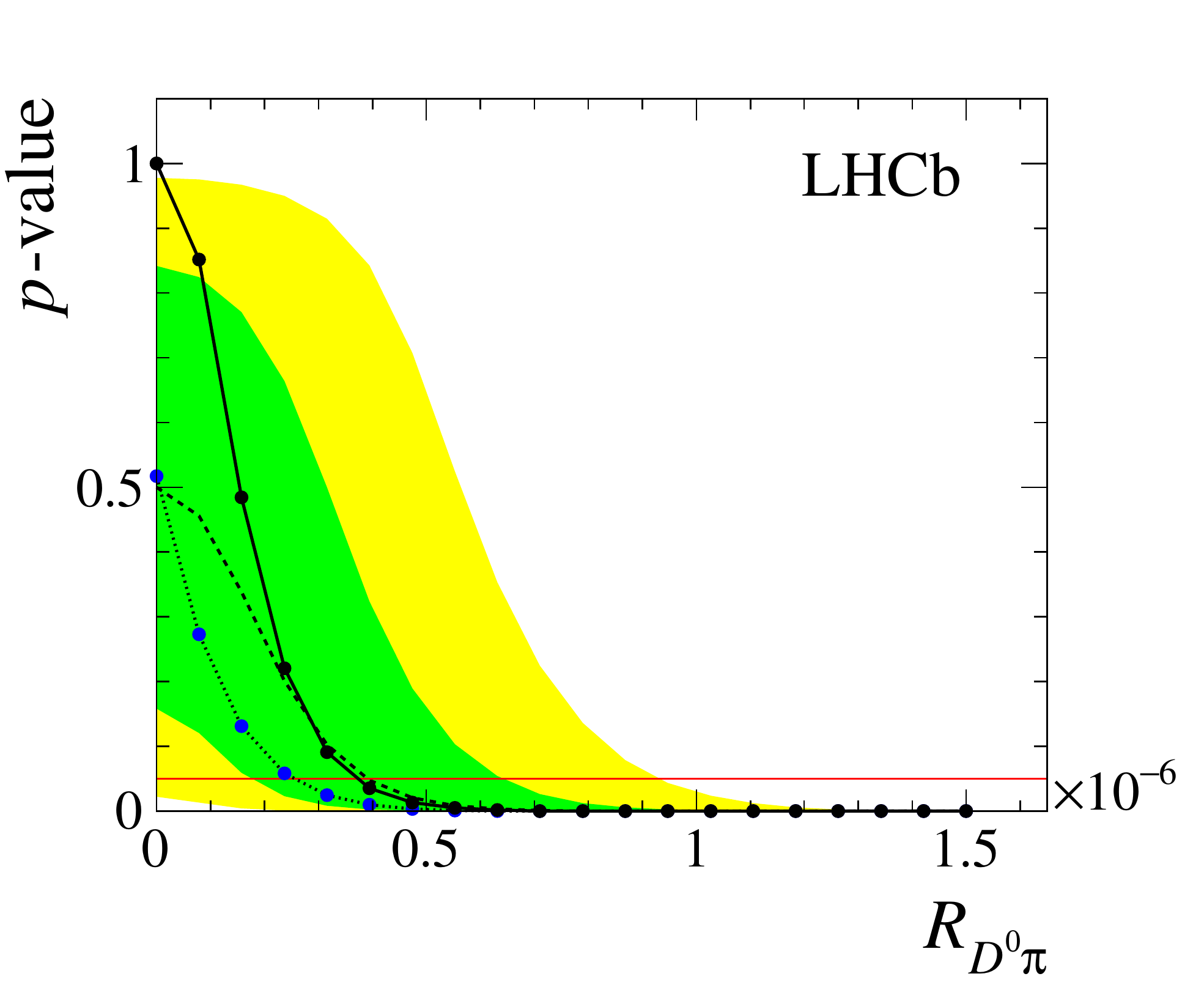}
     \includegraphics[width=0.4\linewidth]{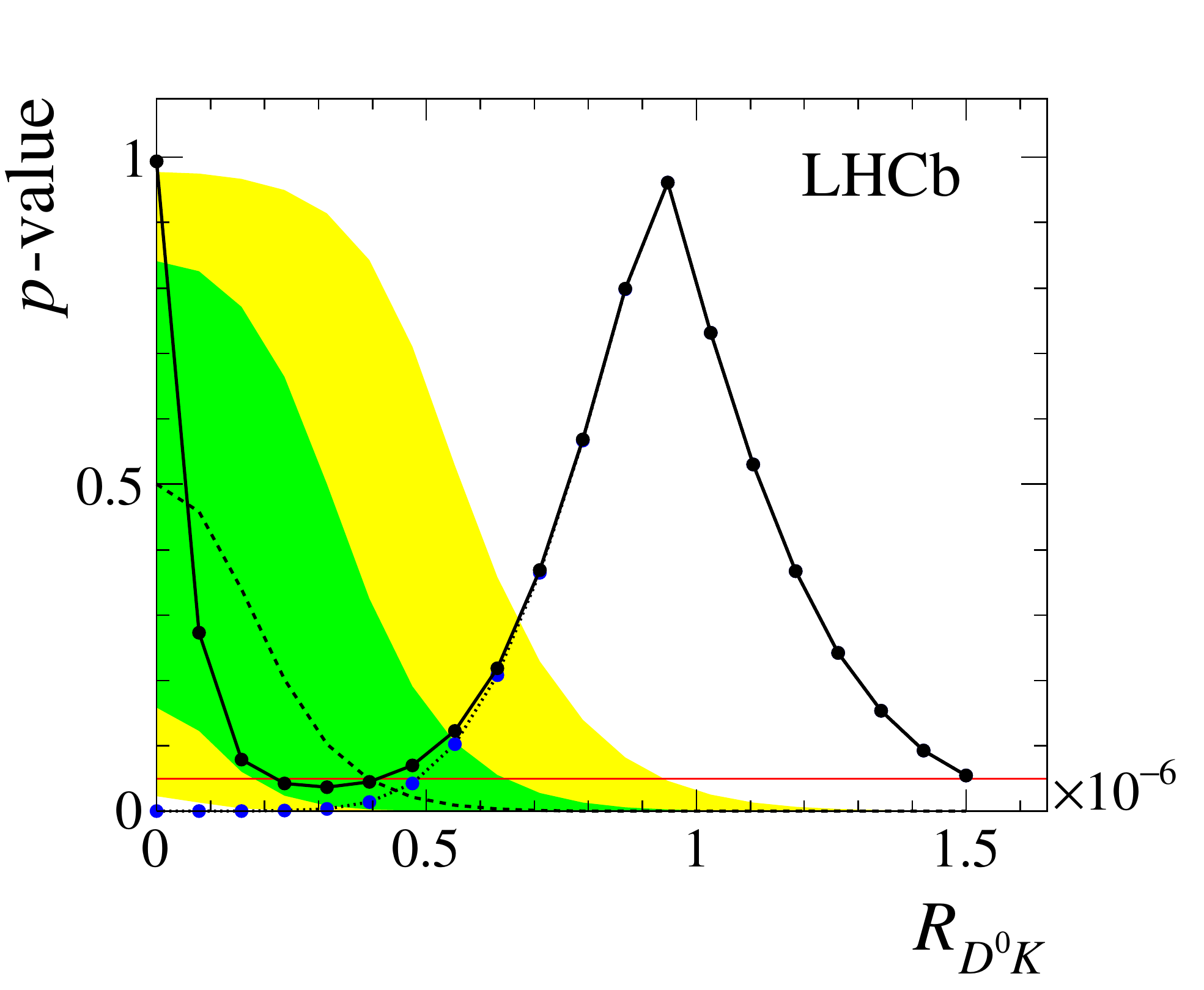} \\
  \end{center}
  \caption{CL$_{\text{s}}$ $p$-value distributions for the $R_{\Dz h}$ observables. The dashed line represents the expected CL$_{\text{s}}$ values, where the $1\sigma$ and $2\sigma$ contours are indicated by the green (dark) and yellow (light) shaded regions, respectively. Upper limits are determined by the points at which the observed $\text{CL}_{\text{s+b}}$ $p$-values (black points connected by straight lines) fall below 5\% (red solid line). Also displayed are the corresponding $\text{CL}_{\text{s}} = \text{CL}_{\text{s+b}} / \text{CL}_{\text{b}}$ values (blue points connected by straight dotted lines). 
  \label{fig:CLs_Plots}}
\end{figure}

To test the significance of each signal yield, $\text{CL}_{\text{s}}$ hypothesis tests~\cite{2010acat.confE..57M} are performed. Upper limits at 95\% confidence level (CL) are determined by the point at which the $p$-value falls below 5\%. All free variables in the fit are considered as nuisance parameters in this procedure. The $p$-value distributions for each $R_{\Dz h}$ measurement are shown in Fig.\,\ref{fig:CLs_Plots}. The $\Bcp \to D^{*0} h^+$ modes demonstrate no excess, and the $R_{D^{*0}h}$ CL$_\text{s}$ confidence intervals are determined similarly to that of $R_{\Dz\pi}$. The upper limits at 95\% confidence level found for $R_{\Dz\pi}$, $R_{D^{*0}\pi}$ and $R_{D^{*0}K}$ are
\begin{align*}
R_{\Dz\pi} &< 3.9 \times 10^{-7}\,, \\
R_{D^{*0}\pi} &< 1.1 \times 10^{-6}\,, \\
R_{D^{*0}K} &< 1.1 \times 10^{-6}. 
\end{align*}
The systematic uncertainties affecting the measurements are found to be much smaller than the statistical uncertainty, and do not alter the above upper limits. 

In the case of $R_{\Dz K}$, the observed signal is of much higher significance. To determine the full uncertainty for $R_{\Dz K}$, the systematic uncertainties affecting the measurement are accounted for. A systematic uncertainty of $1.1 \times 10^{-8}$ is incurred from the use of fixed terms in the invariant mass fit. According to Eq.~(\ref{r_Bc_h_definition}), several terms with associated relative uncertainties scale the measured signal yield: $\xi$ with 5.3\% uncertainty, \mbox{$\mathcal{B}(\Bp \to \Dzb \pip)$ with 3.1\% uncertainty~\cite{Olive:2016xmw}}, and $\mathcal{N}(\Bp \to \Dzb \pip)$ with 0.14\% uncertainty. The total systematic uncertainty, given by the sum in quadrature, is 6.2\%. 

To determine the significance of the $\Bcp \to \Dz\Kp$ peak, a likelihood scan is performed. The resulting $-\Delta\text{log}(L)$ value for the $R_{\Dz K} = 0$ hypothesis corresponds to a statistical significance of \mbox{$\sqrt{-2\Delta\text{log}(L)} = 5.1 \sigma$} for the signal. The final result is
\begin{align*}
R_{\Dz K} &= (9.3\,^{+2.8}_{-2.5} \pm 0.6) \times 10^{-7}\,,
\end{align*}
where the first uncertainty is statistical and the second is systematic. This is the first observation of the $\Bcp \to \Dz \Kp$ decay. The value of $R_{\Dz K}$ is at the high end of theoretical predictions~\cite{Bc_theory,Bc_theory3,Zhang2009} and an expectation based on the observed $\Bcp \to J/\psi \pip$ yield at LHCb~\cite{LHCb-PAPER-2015-024}. From Refs.~\cite{LHCb-PAPER-2014-050} and~\cite{Olive:2016xmw}, $R_{J/\psi\pi} = (7.0 \pm 0.3) \times 10^{-6}$ is obtained. As $f_{c}/f_{u}$ is common to both $R_{J/\psi\pi}$ and $R_{\Dz K}$, the ratio of branching fractions is measured to be \mbox{$\mathcal{B}(\Bc \to \Dz \Kp)/\mathcal{B}(\Bc \to J/\psi \pip) = 0.13 \pm 0.04 \pm 0.01 \pm 0.01$}, where the first uncertainty is statistical, the second is systematic, and the third comes from $R_{J/\psi\pi}$.

The absence of the $\Bcp \to \Dz \pip$ mode shows that the $\Bcp \to \Dz \Kp$ amplitude is not dominated by the tree-level $b \to u$ transition shown in Fig.\,\ref{fig:feyn}(a), but rather by the penguin (b) and/or weak annihilation (c) diagrams. This result constitutes the first observation of such amplitudes in the decay of a $\Bcp$ meson. 


\section*{Acknowledgements}

\noindent We express our gratitude to our colleagues in the CERN
accelerator departments for the excellent performance of the LHC. We
thank the technical and administrative staff at the LHCb
institutes. We acknowledge support from CERN and from the national
agencies: CAPES, CNPq, FAPERJ and FINEP (Brazil); NSFC (China);
CNRS/IN2P3 (France); BMBF, DFG and MPG (Germany); INFN (Italy); 
FOM and NWO (The Netherlands); MNiSW and NCN (Poland); MEN/IFA (Romania); 
MinES and FASO (Russia); MinECo (Spain); SNSF and SER (Switzerland); 
NASU (Ukraine); STFC (United Kingdom); NSF (USA).
We acknowledge the computing resources that are provided by CERN, IN2P3 (France), KIT and DESY (Germany), INFN (Italy), SURF (The Netherlands), PIC (Spain), GridPP (United Kingdom), RRCKI and Yandex LLC (Russia), CSCS (Switzerland), IFIN-HH (Romania), CBPF (Brazil), PL-GRID (Poland) and OSC (USA). We are indebted to the communities behind the multiple open 
source software packages on which we depend.
Individual groups or members have received support from AvH Foundation (Germany),
EPLANET, Marie Sk\l{}odowska-Curie Actions and ERC (European Union), 
Conseil G\'{e}n\'{e}ral de Haute-Savoie, Labex ENIGMASS and OCEVU, 
R\'{e}gion Auvergne (France), RFBR and Yandex LLC (Russia), GVA, XuntaGal and GENCAT (Spain), Herchel Smith Fund, The Royal Society, Royal Commission for the Exhibition of 1851 and the Leverhulme Trust (United Kingdom).


\addcontentsline{toc}{section}{References}
\setboolean{inbibliography}{true}

\ifx\mcitethebibliography\mciteundefinedmacro
\PackageError{LHCb.bst}{mciteplus.sty has not been loaded}
{This bibstyle requires the use of the mciteplus package.}\fi
\providecommand{\href}[2]{#2}

\newpage

\centerline{\large\bf LHCb collaboration}
\begin{flushleft}
\small
R.~Aaij$^{40}$,
B.~Adeva$^{39}$,
M.~Adinolfi$^{48}$,
Z.~Ajaltouni$^{5}$,
S.~Akar$^{59}$,
J.~Albrecht$^{10}$,
F.~Alessio$^{40}$,
M.~Alexander$^{53}$,
S.~Ali$^{43}$,
G.~Alkhazov$^{31}$
% ...

{\footnotesize \it
% ...
$ ^{5}$Clermont Universit{\'e}, Universit{\'e} Blaise Pascal, CNRS/IN2P3, LPC, Clermont-Ferrand, France\\
% ...
$ ^{10}$Fakult{\"a}t Physik, Technische Universit{\"a}t Dortmund, Dortmund, Germany\\
% ...
$ ^{31}$Petersburg Nuclear Physics Institute (PNPI), Gatchina, Russia\\
% ...
$ ^{39}$Universidad de Santiago de Compostela, Santiago de Compostela, Spain\\
$ ^{40}$European Organization for Nuclear Research (CERN), Geneva, Switzerland\\
% ...
$ ^{43}$Nikhef National Institute for Subatomic Physics, Amsterdam, The Netherlands\\
% ...
$ ^{48}$H.H. Wills Physics Laboratory, University of Bristol, Bristol, United Kingdom\\
% ...
$ ^{53}$School of Physics and Astronomy, University of Glasgow, Glasgow, United Kingdom\\
% ...
$ ^{59}$University of Cincinnati, Cincinnati, OH, United States\\
% ...
}
\end{flushleft}

\end{document}